\documentclass[pra,twocolumn,letterpaper,showpacs,amsmath,amssymb,floatfix]{revtex4}
\usepackage[dvips]{graphicx}
\usepackage{amssymb}
\begin{document}

\newcommand{\ket}[1]{|#1\rangle}
\newcommand{\bra}[1]{\langle #1|}
\newcommand{\bracket}[2]{\langle #1|#2\rangle}
\newcommand{\ketbra}[1]{|#1\rangle\langle #1|}
\newcommand{\average}[1]{\langle #1\rangle}
\newtheorem{theorem}{Theorem}

\title{Spherical Code Key Distribution Protocols for Qubits}
\author{Joseph M. Renes}
\affiliation{Department of Physics and Astronomy, University of New Mexico,\\
Albuquerque, New Mexico 87131--1156, USA\\
\texttt{renes@phys.unm.edu}}

\begin{abstract}
Recently spherical codes were introduced as potentially more capable
ensembles for quantum key distribution. Here we develop specific key
creation protocols for the two qubit-based spherical codes, the trine
and tetrahedron, and analyze them in the context of a suitably-tailored 
intercept/resend attack, both in standard form, and a ``gentler'' version whose
back-action on the quantum state is weaker. 
When compared to the standard unbiased basis
protocols, BB84 and six-state, two distinct advantages are found.
First, they offer improved tolerance of eavesdropping, the trine besting
its counterpart BB84 and the tetrahedron the six-state protocol. 
Second, the key error rate
may be computed from the sift rate of the protocol itself, removing
the need to sacrifice key bits for this purpose. This simplifies the
protocol and improves the overall key rate.
\end{abstract}

\pacs{03.67.Dd, 03.67.Hk, 03.67.-a}

\maketitle

Heretofore quantum key distribution protocols have often been
constructed using sets of unbiased bases, enabling key bit creation
whenever the two parties Alice and Bob happen to send and measure the
quantum system in the same basis. Alice randomly selects a basis and a
state within that basis to send to Bob, who randomly chooses a basis in
which to measure and decodes the bit according to their pre-established
scheme. Should Bob choose the same basis as Alice, his outcome is
perfectly correlated with hers. Each of the parties publicly announces
the bases used, and for each instance they agree, they establish one
letter of the key. The use of more than one basis prevents any would-be
eavesdropper Eve from simply reading the encoded bit without Alice and
Bob noticing. In two dimensions two sets of mutually unbiased bases
exist, forming the BB84~\cite{bb84} and six-state
protocols~\cite{bruss98}.

Equiangular spherical codes can be used to construct a new scheme for
key distribution~\cite{renes03}. Two such codes exist in two dimensions.
In the Bloch-sphere representation we may picture these ensembles as
three equally-spaced coplanar states forming a trine or four
equally-spaced states forming a tetrahedron. Both Alice and Bob replace
their use of unbiased bases with equiangular spherical codes; by
arranging Bob's code to be the dual of Alice's, key creation becomes a
process of elimination, as previously considered by 
Phoenix, \emph{et al.}~\cite{pbc00}. Each of Bob's measurement
outcomes is orthogonal to one of Alice's signals, and thus each outcome excludes
one signal. 
Alice may then attempt to furnish the remaining information by announcing 
a certain number of signals that were not sent, a process known as sifting.
By symmetry, Bob can also send the sifting information to Alice, in the form 
of outcomes not obtained; this convention will be followed here.
The shared (anti-) correlation
between signal and outcome allows them to remain one step ahead of an
eavesdropper Eve, ensuring that unless she tampers with the quantum
signal, she knows nothing of the created key.

Should Eve tamper with the signal, the disturbance can be recognized by Alice and Bob in the
statistics of their results. With this they can determine what she
knows about their key, and they may either proceed to shorten their key
string so as to remove Eve's information of it, or else discard it
entirely and begin anew. Unlike bases-based protocols, however, here
Alice and Bob can determine the disturbance from the sifting rate directly,
obviating the need to explicitly compare (and waste) portions of the key for this purpose.

The overarching questions in evaluating a key
distribution protocol are whether or not it is unconditionally secure,
and if so, what the maximum tolerable error rate is. If, by granting
Eve the ability to do anything consistent with the laws of physics,
Alice and Bob can still share a key, the protocol is said to be
unconditionally secure. This state of affairs persists up to the
maximum tolerable error rate, at which point Alice and Bob must abandon
their key creation efforts. Establishing unconditional security
is complicated and delicate, so here we restrict attention to more
limited attacks, examining the intercept/resend attack and a
``gentler'' variant. In these settings we find that
the spherical codes are more tolerant of noise than their basic 
counterparts.
First, however, we must consider the protocols for the two spherical
codes in detail.

Unlike the case of unbiased bases, in which Alice's choice of signal or
Bob's outcome determines the key letter, for the trine and tetrahedron
it is only the relation between Alice's signal and Bob's outcome that
determines the bit. In the trine protocol Alice's choice of signal
narrows Bob's possible outcomes to the two lying 60 degrees on either
side. Each is equally likely, and they publicly agree beforehand that
the one clockwise from Alice's signal corresponds to \texttt{1} and the other
\texttt{0}. Alice hopes to determine which is the case when Bob announces one
outcome that he \emph{did not} receive. For any given outcome, he
chooses randomly between the other two and publicly announces it. Half
the time he announces that he did not receive the outcome which Alice
already knew to be impossible. This tells Alice nothing new, and she
announces that the protocol failed. In the other half of cases, Alice
learns Bob's outcome and announces success. 

Upon hearing his message was a success, Bob can determine the signal
Alice sent. For any outcome Bob receives, he immediately knows one
signal Alice could not have sent, and the message that his announcement
was successful indicates to him that she also did not send the signal
orthogonal to his message. Had she sent that signal, she would have
announced failure; thus Bob learns the identity of Alice's signal. Each
knowing the relative position of signal and outcome, they can each
generate the same requisite bit. This round of communication is the analog 
of sifting in the protocols utilizing unbiased bases: a follow-up 
classical communication referencing the quantum signals with which 
Alice and Bob establish the key.

Mathematically, we might consider the protocol as follows. Alice sends
signal $j$, and Bob necessarily obtains $k=j\!+\!1$ or $k=j\!+\!2$. He
announces that he did not receive $l\!\neq\! k$. If $l\!=\!j$, Alice
announces failure. Otherwise each party knows the identity of $j,k$,
and $l$, and they compute the key bit as $ (1\!-\!\epsilon_{jkl})/2$.
Fig.~1 shows the case that they agree on a \texttt{1}.

\begin{figure}[h]
\label{fig:label}
\includegraphics[scale=.6]{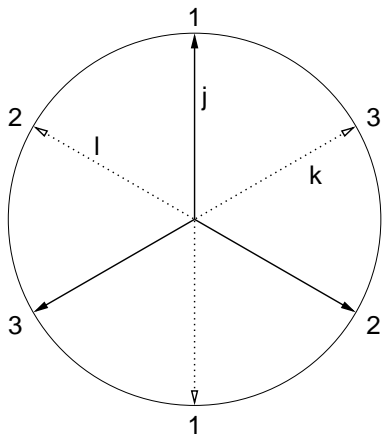}
\caption{Bloch-sphere representation of the trine-based protocol by
which Alice and Bob create a secret key bit, shown here creating a \texttt{1}.
Alice's three possible signal states are shown in black and Bob's
measurement outcomes in dotted lines; antipodal points are orthogonal.
Without loss of generality we may assume that Alice sends the state
$j=1$. The antipodal point is the impossible outcome for Bob; here he
obtains the outcome $k=3$. Of the two outcomes he did not get, he picks
one at random and announces this to Alice. Here he announces the
outcome $l=2$, and Alice infers the value of $k$. Had Bob announced the
other outcome, the protocol would fail, as this tells Alice
nothing she does not already know. Here she
announces that she is satisfied with Bob's message, and Bob infers the
value of $j$, since Alice's signal could not have been $l$. Now they
compute the bit $(1\!-\!\epsilon_{jkl})/2=1$. The announcement only
reveals $l$, so the bit is completely secret.}
\end{figure}

Though Eve may listen to the messages on the classical channel, she
still has no knowledge of the bit value, for all she knows is one
outcome Bob did not receive and the corresponding antipodal state not 
sent by Alice. Of the two remaining equally-likely alternatives,
one corresponds to a \texttt{0} and the other a \texttt{1}. Hence the protocol
establishes one fully secret bit half the time, analogous to the BB84
protocol.

The strategy for the tetrahedron is entirely similar, except that Bob
must now reveal two outcomes not obtained. As depicted in Fig.~2,
Alice uses four tetrahedral states in the Bloch-sphere picture, and as
before Bob uses the dual of Alice's tetrahedron for measurement. Alice
sends signal $j$ and Bob receives $k\!\neq\! j$. He then randomly chooses
two outcomes $l$ and $m$ he did not obtain and announces them.
One-third of the time this is successful, in that $l\!\neq\! j$ and $m\!\neq\!
j$. This allows Alice to infer $k$, and her message of satisfaction
allows Bob to infer $j$, just as for the trine.  They then each compute
the bit $(1+\epsilon_{jklm})/2$.

Again they stay one step ahead of Eve as she listens to the messages,
as she can only narrow Alice's signal down to two possibilities. Given
the order of Bob's messages, one of these corresponds to \texttt{0} and the
other to \texttt{1}, so Eve is ignorant of the bit's identity. Using the
tetrahedron allows Alice and Bob to establish one fully secret bit one
third of the time, analogous to the six-state protocol.

\begin{figure}
\label{fig:tetra}
\includegraphics[scale=.7]{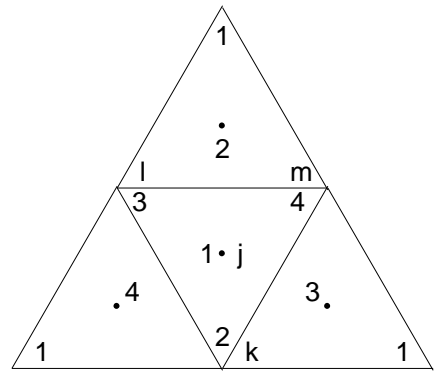}
\caption{Unfolded view of the Bloch-sphere tetrahedron states. Vertices
of triangles correspond to Bob's outcomes, their centers Alice's
signals; all three vertices of the large triangle represent the same
point antipodal to its center. Suppose Alice sends signal $j$; Bob
necessarily receives $k\neq j$. Here we suppose $j=1$ and $k=2$. Bob
then announces two outcomes not obtained, here shown as $l=3$ and
$m=4$. Had either message equaled $j$, which happens 2/3 of the time,
Alice announces failure. Otherwise, as here, she accepts.  Thus Alice
determines $k$, and Bob finds out $j$.  They compute the bit
$(1+\epsilon_{jklm})/2=1$. The announcement reveals only $l$ and $m$,
so the bit is secret.}
\end{figure}

In the two protocols, the dual arrangement of signals and measurements
allows Alice and Bob to proceed by elimination to establish a putative
key. To ensure security of the protocols, however, the arrangement must
also disallow Eve from reading the signal without Alice and Bob
noticing. Analyzing the intercept/resend attack provides evidence of
how well the protocols based on spherical codes measure up to this
task.

If Eve tampers with the signals in order to learn their identity, the
inevitable disturbance allows Alice and Bob to infer how much Eve knows
about the raw key. They can then proceed to use error correction and
privacy amplification procedures to distill a shorter key which, with
high probability, is identical for Alice and Bob and which Eve has low
probability of knowing anything about. Instead of delving into the
details of error correction and privacy amplification, we may instead
use a lower bound on the optimal rate of the distilled key, i.e., its
length as a fraction of the raw key.~\cite{ck78} This provides a
reasonable guess as to what may be achieved in practice and is known to
be achieveable using one-way communication. Given $N\!\rightarrow\!\infty$
samples from a tripartite distribution $p(a,b,e)$, Alice and Bob can 
construct a protocol to distill with high
probability a length $RN$ string about which
Eve has asymptotically zero information for
\begin{equation}
\label{eq:ratebound}
R= I(A\!:\!B)-\min\{I(A\!:\!E),I(B\!:\!E)\}.
\end{equation}
Here the tripartite distribution refers to Alice's and Bob's
bit values $a$ and $b$, and Eve's best guess $e$ from the
eavesdropping. The quantity $I(A\!:\!B)$ is the mutual information
between two parties, quantifying how much knowledge of 
one's outcome implies about the other's. 

Here we're assuming that Eve simply intercepts a fraction $q$ of the
signals, measures them, and sends a new state on to Bob. The first task
is then to determine $R$ as a function of $q$ and then to relate $q$ to
the statistics compiled in the course of the protocol. As it happens, Eve's best
attack in the intercept/resend context 
is to use \emph{both} Alice's and Bob's trines for measurement, half
the time pretending to be Alice and the other half Bob. This holds for
the tetrahedron as well and is due to the minimum in
Eq.~(\ref{eq:ratebound}), which gives the equation a symmetry between Alice and 
Bob with respect to Eve. Choosing only one of the trines (or tetrahedra) to measure
breaks this symmetry, leading the minimum to pick the smaller information quantity and 
yield a consequently larger key. 
By mixing the two strategies, Eve restores the symmetry and increases the minimum knowledge she has
about either party's bit string. Phoenix, \emph{et al}.~\cite{pbc00} note that the scheme in which 
Eve pretends to be Bob maximizes
her mutual information with Alice; however, as the analysis stops there and 
does not proceed to consider either Eve's information about the key bits nor 
any secret key rate bounds, it is insufficient as a cryptographic analysis.
 
To determine the mutual information quantities as functions of $q$, it
suffices to consider first the case in which Eve intercepts every
signal and uses Alice's ensemble for measurement. With these quantities
in hand, we can mix Eve's two measurement strategies appropriately and then include her
probability of interception. We begin with the trine. Given a signal
state from Alice, there are two cases to consider. Either Eve measures
and gets the same state, which happens with probability 2/3, or she
obtains one of the other two results, with probability 1/6 for each.
Whatever her outcome, she passes the corresponding state along to Bob
and guesses that it was the state sent by Alice, \emph{unless the
subsequent exchange of classical messages eliminates this possibility},
at which point she reserves judgement about the key bit.

Suppose Eve's outcome corresponds to Alice's signal, and thus no
disturbance is caused. Naturally, Alice and Bob go on to establish a
bit half the time, a bit known to Eve. On the other hand,
should her outcome not coincide with Alice's signal, there are two
further possibilities. Half the time Bob obtains a result consistent
with Alice's signal, i.e. not the orthogonal state, 
and a further half the time the sifting succeeds.
However, the required sifiting messages will
eliminate Eve's outcome as Alice's signal, thus forcing Eve to abandon 
her guess. 
In the remaining case, Bob's result is orthogonal to Alice's signal, which 
guarantees successful sifting, but also different bit values for Alice and Bob.
Half of Eve's guesses are excluded while the remainder agree with Bob's.

Putting all this together, one obtains that the protocol fails with
probability 5/12. All three agree one-third of the time, and Alice's bit is
different from that shared by Bob and Eve one-twelfth of the time. In the 
remaining one-sixth of events, Eve does not field a guess, as the messages exchanged
by Alice and Bob contradict her measurement results; better to abstain 
than to introduce a purely random guess. In this subset of events, Alice and 
Bob agree a further half the time. 

Of the key bits created, Bob and Alice agree with
probability 5/7, while Eve and Alice agree
with probability 4/7. Eve only fields a guess with probability 5/7, and
always agrees with Bob when she does.
These numbers are obtained by considering the raw
probabilities of agreement and renormalizing by $12/7$. 
Should Eve
instead measure the signals using Bob's trine ensemble, her agreement
probabilities with Alice and Bob are swapped. Mixing the two
strategies then yields Eve a no-guess probability of 2/7,
an agreement probability with either party of 9/14, and an 
an error probability of 1/14.

To interpolate between the endpoints of no interception and full
interception, note that to condition on the cases of successful bit
creation, the probability of bit agreement must be renormalized by the
probability of sifting success. This probability depends linearly on $q$:
$p_{\rm sift}=(6+q)/12$. All probabilities must therefore contain
$6+q$ in the denominator, whence we may derive pairwise probabilities that the parties' bit values
agree:
$p_{ab}=(6-q)/(6+q)$, and $p_{ae}=9q/2(6+q)$, respectively.
Eve's probability to not guess at all is $2(3-2q)/(6+q)$. Determining
the relevant mutual informations from these expressions is
straightforward; for expressions involving Eve, simply treat the ``no-guess'' as another 
outcome which has no correlation at all to the other party.

By determining the probability of error in Alice's and Bob's bit
strings as a function of $q$, we may compare to other protocols. For
the trine, errors occur in the key string with probability $2q/(6+q)$.
Using the calculated agreement probabilities 
in the rate bound,
one obtains that $R=0$ corresponds to a maximum
tolerable bit error rate of 20.4\%. This compares favorably with the BB84
protocol's maximum tolerable bit error rate of 17.1\% 
under the same attack~\cite{ekerthuttner94}.
In terms of \emph{channel} error rate these figures double, if we consider
the quantum channel to be a depolarizing channel instead of arising
from Eve's interference. If Bob receives the
maximally-mixed state instead of Alice's signal, the probability of error 
given successful sifting is 1/2. Hence
a fully depolarizing channel leads to a bit error rate of 50\% for either 
protocol. 

Analysis of the tetrahedron protocol proceeds similarly by examining
the various cases. In this case, when $q=1$ the failure rate of the
protocol drops to 5/9, while Alice and Bob agree with probability 5/8,
Eve has probability 7/16 of knowing Alice's or Bob's bit value, and she 
reserves judgement half the time. As
the successful sifting rate of the protocol goes like $(3+q)/9$, we may determine
the form of the probabilities using the same method to be
$p_{ab}=p_{eb}=(6-q)/2(3+q)$ and $p_{ae}=7q/4(3+q)$,
while the error rate in the key string is $3q/2(3+q)$ and Eve's probability 
of not guessing is $(3\!-\!q)/(3\!+\!q)$. Using
these probabilities in the rate bound yields a maximum error rate of
26.7\%. Like before,
this compares favorably to the maximum tolerable error rate in the
six-state protocol of 22.7\%. 

Eve's attack could be gentler, however. In the version already considered, her POVM consists of subnormalized
projectors onto the code states in addition to an element proportional
to the identity operator, corresponding to the case in which Eve
opts not to intercept the signal. A similar POVM can be created by distributing a piece of identity operator
to all the other elements.
The crucial difference is that the state Eve sends on to Bob after
her measurement is different.  Using the square root of each POVM
element in the formula for the post-measurement state,
the resulting measurement yields
Eve more information for the same amount of disturbance. Note that in the context
of the BB84 protocol, this attack was determined to be optimal when Eve
does not wait to hear in which basis the signal was
prepared~\cite{luetkenhaus96}. 

Enlisting the aid of Mathematica to carry out the bookkeeping yields 
the following results. Since the attack is stronger, the maximum tolerable error decreases; 
in particular the trine can create secret keys up to a 16.6\% bit error rate, as opposed to 15.3\% for its cousin BB84. 
The tetrahedron remains the most robust, sustaining key creation up to a maximum error rate of 22.6\%,
as compared to 21.0\% for the six-state protocol. 

Framing the key rate in terms of the error rate is solely for ease of
comparison, as it is not necessary for Alice and Bob to sacrifice key
bits in order to obtain an estimate of $q$ when using spherical codes,
in contrast to the situation for the unbiased bases. For spherical
codes, the sifting rate of the protocol itself determines $q$; as the
channel becomes noisier and Bob's outcome less correlated with Alice's
signal, the sifting rate increases. Of course, not all of this increase
provides useful key: most of it leads to errors. But Eve cannot
substitute signals solely for the purpose of modifying the sift
rate, as her signals will be uncorrelated with Alice's and will
therefore also lead to an increase in the sift rate. Hence she is
precluded from masking her interceptions, and Alice and Bob can
determine $q$ from the sifting rate itself.

Finally, a word on the feasibility of implementing such protocols. 
Generation
of trine or tetrahedral codewords as polarization states of (near) single-photon
sources is not difficult. 
The generalized measurements accompanying the ensembles can be
performed by using polarizing beam splitters and wave plates to map
polarization states into different propagation modes and proceeding from
there with linear optical elements to produce the appropriate
interference.
Such measurements have indeed been 
performed with rms errors in observed statistical distributions of a few 
percent~\cite{ckcbrs01}. The physical implementation needn't be identical
to the logical construction of the protocol, however. For instance, three
states constructed from two pairs of singlets together with ordinary photodectors
can implement the trine protocol~\cite{bglps04}.


Two advantages 
of using spherical codes have been established herein. First and foremost is the strong
possibility of improved eavesdropping resistance. Subsequent analyses either of
stronger attacks, such as use of an asymmetric cloning machine~\cite{cerf00}, or
the use of error-correcting codes to beat back noise~\cite{shorpreskill00} are required to 
demonstrate this fact in the setting of unconditional security,
though the intercept/resend attacks are indicative of the
trend~\cite{fggnp97}. Beyond security is the ability to directly estimate the
error rate from the sift rate itself, obviating any need to sacrifice raw key bits.

The author acknowledges helpful input from D.~Bru\ss, C.~M.~Caves,
J.~Eisert, D.~Gottesman, and N.~L\"utkenhaus. This work was supported in part by
Office of Naval Research Grant No.~N00014-00-1-0578.

\end{document}